\documentclass[aps,prd,groupaddress,floatfix,tighten,nofootinbib,showpacs,amsfonts,superscriptaddress]{revtex4-2}
\usepackage{url}
\usepackage{xcolor}
\usepackage{hyperref}
\usepackage[latin1]{inputenc}

\colorlet{shadecolor}{gray!15}
\definecolor{greenLinks}{rgb}{0, 0.6, 0}
\definecolor{blueLinks}{rgb}{0, 0, 0.6}
\definecolor{redLinks}{rgb}{0.6, 0, 0}
\definecolor{tempText}{rgb}{0.55, 0.10,0.67}
\definecolor{eprintLinks}{rgb}{0.4, 0.4, 0.4}
\definecolor{journalLinks}{rgb}{0.6, 0, 0}
\usepackage{amssymb}
\usepackage{amsmath,color}
\usepackage{epsfig}
\usepackage{graphicx,epsf,epsfig}
\usepackage{footnote}
\usepackage{multirow}
\usepackage{enumitem}
\usepackage{color}
\def\slc#1{\setbox0=\hbox{$#1$}                  
    \dimen0=\wd0                                 
    \setbox1=\hbox{/} \dimen1=\wd1               
    \ifdim\dimen0>\dimen1                        
       \rlap{\hbox to \dimen0{\hfil/\hfil}}      
       #1                                        
    \else                                        
       \rlap{\hbox to \dimen1{\hfil$#1$\hfil}}   
       /                                         
    \fi}

\def\be{\begin{equation}}
\def\ee{\end{equation}}
\def\gs{\mathrel{
   \rlap{\raise 0.511ex \hbox{$>$}}{\lower 0.511ex \hbox{$\sim$}}}}
\def\ls{\mathrel{
   \rlap{\raise 0.511ex \hbox{$<$}}{\lower 0.511ex \hbox{$\sim$}}}}

\newcommand{\ba}{\begin{array}{c}}
\newcommand{\baz}{\begin{array}{cc}}
\newcommand{\barrr}{\begin{array}{rrr}}
\newcommand{\bad}{\begin{array}{ccc}}
\newcommand{\bav}{\begin{array}{cccc}}
\newcommand{\baf}{\begin{array}{ccccc}}
\newcommand{\bea}{\begin{equation} \begin{array}{c}}
\newcommand{\eea}{\end{array} \end{equation}}
\newcommand{\ea}{\end{array}}

\usepackage[normalem]{ulem}

\def\21{$\mathrm{SU(2)_L \otimes U(1)_Y}$ }

\newcommand {\ignore}[1]{}

\newcommand{\vt}{\vert}
\newcommand{\nn}{\nonumber}

\allowdisplaybreaks \allowdisplaybreaks[2]
  
\newcommand{\AddrCECYT}{Centro de Estudios Cient\'ificos y Tecnol\'ogicos No 16, Instituto Polit\'ecnico Nacional, Pachuca: Ciudad del Conocimiento y la Cultura, Carretera Pachuca Actopan km 1+500, San Agust\'in Tlaxiaca, Hidalgo, M\'exico.\\ 
}

\begin{document}
\title{
Flavored multiscalar $\mathbf{S}_{3}$ model with normal hierarchy neutrino mass} 
%
\author{J. D. Garc\'ia-Aguilar}
\email{jdgarcia@ipn.mx}
\affiliation{\AddrCECYT}

\author{Juan Carlos G\'omez-Izquierdo}
\email{cizquierdo@ipn.mx\\
}
\affiliation{\AddrCECYT}
%

%

%

\date{\bf \today} 

\begin{abstract}\vspace{2cm}
We construct a multiscalar and non-renormalizable model where the $\mathbf{S}_{3}$ flavor symmetry drives mainly the Yukawa couplings. In the quark sector, the Nearest Neighbor Interaction (NNI) textures are behind the CKM mixing matrix so that this is fitted in good agreement with the last available results. In the lepton sector, an almost diagonal charged lepton mass matrix and extended Fritzsch mass textures in the effective neutrino mass matrix, that comes from the type II see-saw mechanism, provide consistent values for the mixing angles where the normal hierarchy is favored. The model predicts a CP-violating phase consistent with the experimental data and the BR for the lepton flavor violation process, $\mu\rightarrow e\gamma$, is well below the current bound.

\end{abstract}

%
\maketitle
%

\section{Introduction}

Although the Standard Model (SM) describes successfully the fundamental interactions among the particles, it is clear that this has to be extended to incorporate the neutrino mass and its mixings. Along with this, the flavor puzzle remains unsolved.

As it is well known, the CKM matrix ~\cite{Cabibbo:1963yz, Kobayashi:1973fv} is almost the identity one which differs completely to the PMNS matrix ~\cite{Maki:1962mu, Pontecorvo:1967fh} where large values, in its entries, can be found. The pronounced hierarchy among the quark masses,
$m_{t}\gg m_{c}\gg m_{u}$ and $m_{b}\gg m_{s}\gg m_{d}$, could explain the small mixing angles, that parametrize the CKM, since that those
depend strongly on the mass ratios \cite{Fritzsch:1999ee,Xing:2014sja,Verma:2015mgd}.
This notable hierarchy may naturally come from the hierarchical mass matrices, as for example, the Fritzsch~ \cite{Fritzsch:1977za, Fritzsch:1977vd, Fritzsch:1979zq, Fritzsch:1985eg,Fritzsch:1999ee,Fritzsch:1999rb, Fritzsch:2015foa} and nearest neighbor interaction (NNI)~\cite{Branco:1988iq, Branco:1994jx, Harayama:1996am, Harayama:1996jr} mass textures. Although both textures give us the extended Gatto-Sartori-Tonin relations
\cite{Gatto:1968ss,Cabibbo:1968vn,Oakes:1969vm,Fritzsch:1977za,Fritzsch:1977vd,Fritzsch:1979zq,Fritzsch:1999ee,Fritzsch:1999rb}, the former one
presents some problems~\cite{Branco:2010tx, Fritzsch:2011cu} with the top mass and the CKM element,  $V_{cb}$. Actually, the NNI textures are capable of fitting with great accuracy the CKM mixing matrix.

In the lepton sector, the neutrino mass nature is still an open question, several mechanisms have addressed this subtle issue but there is no a conclusive answer yet. Speaking about the hierarchy, this is weak among the lepton masses, at least, in the active neutrinos. 
It has been widely believed that this weak hierarchy is responsible for getting large values for PMNS matrix elements. Nonetheless, the hierarchical Fritzsch textures fit quite well the lepton mixings for the normal ordering in the neutrino masses \cite{Fritzsch:2015foa}. Otherwise, the larges values in the PMNS mixing matrix can be understood elegantly by the presence of some discrete symmetry, in the neutrino mass matrix, which is exhibited in the neutrino mixing. In here, we can find the $\mu\leftrightarrow \tau$ symmetry~\cite{Fukuyama:1997ky, Mohapatra:1998ka, Lam:2001fb, Kitabayashi:2002jd, Grimus:2003kq,Koide:2003rx, Fukuyama:2020swd}, Tri-Bimaximal~\cite{Harrison2002167, Xing200285, Altarelli:2012ss, Rahat:2018sgs, Perez:2019aqq, Rahat:2020mio}, Cobimaximal mixing matrices \cite{Fukuura:1999ze, Miura:2000sx, Ma:2002ce,Ferreira:2016sbb, Ma:2016nkf, Ma:2017moj, Ma:2017trv, Grimus:2017itg,CarcamoHernandez:2017owh,CarcamoHernandez:2018hst,CarcamoHernandez:2020udg}. Currently, a plethora of interesting flavored models exist but some of them will be ruled out by the experimental data~\cite{Esteban:2018azc} which seem favor the normal hierarchy instead of the inverted one, and the same time the Dirac CP-violating phase, $\delta_{CP}$, is close to $270^{\circ}$.

From the model building point of view, a large effort has been done to use the flavor symmetries~\cite{Ishimori:2010au,Grimus:2011fk,Ishimori:2012zz,King:2013eh} to get desirable textures in the fermion mass matrices, and therefore, the well known mixing patterns. In the literature, flavor symmetries
have been applied separately to the quark and lepton sector but the most challenged task has been to apply these symmetries in a unified way to both sectors. In this line of thought, a well known and studied  flavor symmetry is the $\mathbf{S}_{3}$ non-abelian group that has been explored exhaustively in different frameworks~\cite{Pakvasa:1977in, Gerard:1982mm, Kubo:2003iw, Kubo:2003pd,Kobayashi:2003fh, Chen:2004rr, Kubo:2005sr, Felix:2006pn, Mondragon:2007af, Mondragon:2007nk, Mondragon:2007jx, Meloni:2010aw, Dicus:2010iq, Dong:2011vb, Canales:2011ug, Canales:2012ix, Kubo:2012ty, Canales:2012dr, GonzalezCanales:2012kj, Dias:2012bh, GonzalezCanales:2012za, Meloni:2012ci, Canales:2013ura, Ma:2013zca, Canales:2013cga, Hernandez:2014lpa, Hernandez:2014vta, Ma:2014qra, Gupta:2014nba, Hernandez:2015dga, Hernandez:2015zeh, Hernandez:2015hrt, Arbelaez:2016mhg, Hernandez:2013hea, CarcamoHernandez:2016pdu, Das:2014fea, Das:2015sca, Pramanick:2016mdp, Das:2017zrm, Cruz:2017add, Gomez-Izquierdo:2017rxi, Garces:2018nar, Gomez-Izquierdo:2018jrx, Ge:2018ofp, Das:2018rdf, Xing:2019edp, Pramanick:2019oxb, Kuncinas:2020wrn, Vien:2020trr}, and this seems to work quite well for both sectors. In addition, an interesting research on dark matter has been carried out and the $\mathbf{S}_{3}$ flavor symmetry plays an crucial role to tackle this issue \cite{Espinoza:2018itz,Espinoza:2020qyf}. These studios make the $\mathbf{S}_{3}$ discrete group attractive to be the flavor symmetry at low energy, at the same time, this encourages us to look for the best framework where the flavor symmetry solve the majority of the open questions.

In this paper, we take a different route to previous studies~\cite{Kubo:2003iw, Kubo:2003pd, Kubo:2005sr, Felix:2006pn, Mondragon:2007af, Mondragon:2007nk, Mondragon:2007jx, Canales:2011ug, Canales:2012ix, Kubo:2012ty, Canales:2012dr, GonzalezCanales:2012kj,  GonzalezCanales:2012za, Canales:2013ura, Canales:2013cga, Cruz:2017add, Gomez-Izquierdo:2017rxi, Garces:2018nar, Gomez-Izquierdo:2018jrx} on the $\mathbf{S}_{3}$ flavor symmetry where a minimal SM extension, with three right-handed neutrinos, was enlarged by three Higgs doublets to reproduce the CKM and PMNS mixing matrices; in the latter sector, the inverted hierarchy was favored. Therefore, we left out the renormalizable framework and the minimal matter content to build a model where hierarchical mass matrices are behind the CKM and PMNS.
Then, we construct a multiscalar and non-renormalizable model where the $\mathbf{S}_{3}$ flavor symmetry drives mainly the Yukawa couplings. In the quark sector, the Nearest Neighbor Interaction (NNI) textures are behind the CKM mixing matrix so that this is fitted in good agreement with the last available results. In the lepton sector, an almost diagonal charged lepton mass matrix and extended Fritzsch mass textures in the effective neutrino mass matrix, that comes from the type II see-saw mechanism, provide consistent values for the mixing angles where the normal hierarchy is favored. The model predicts a CP-violating phase consistent with the experimental data and the BR for the lepton flavor violation process, $\mu\rightarrow e\gamma$, is well below the current bound.

The plan of the paper is as follows: the model, the matter content and the fermion mass matrices are described in detail in section II; in the section III, the CKM and PMNS mixing matrices are obtained and relevant features are remarked. In addition, an analytical study is carried out on the mixing angles to find the parameter space that accommodates the observables.
A numerical study is carried out, in section IV, and we find a set of values for the free parameters that fit the mixings. We close with relevant conclusions in section V.

\section{The Model}

The current framework is a scalar extension of the SM so that the usual matter content under the gauge group $\mathbf{SU(3)}_{C}\otimes \mathbf{SU(2)}_{L} \otimes \mathbf{U(1)}_{Y}$ is considered. Then, in the Higgs sector we have added two extra Higgs doublets, apart from it, some flavon scalars have been included in the quark and lepton sector to get the desirable mass textures. In this way, under the gauge group, the fermion fields are
\begin{eqnarray}
Q_{L}&=&\begin{pmatrix}
u_{L}\\ d_{L}
\end{pmatrix}\sim \left(3, 2, \frac{1}{3}\right), \qquad d_{R}\sim \left(3, 1,-\frac{2}{3}\right),\qquad u_{R}\sim \left(3, 1,\frac{4}{3}\right);\nn\\
L&=&\begin{pmatrix}
\nu_{L}\\ e_{L}
\end{pmatrix}\sim \left(1, 2,-1\right),\qquad e_{R}\sim \left(1, 1,-2\right)
\end{eqnarray}
Additionally, in the scalar sector 
\begin{equation}\label{scal}
H=\begin{pmatrix}
H^{+}\\ H^{0}
\end{pmatrix}\sim \left(1, 2, 1\right),\quad \Delta= \left( 
\begin{array}{cc}
\frac{\Delta ^{+}}{\sqrt{2}} & \Delta ^{++} \\ 
\Delta^{0} & -\frac{\Delta ^{+}}{\sqrt{2}}%
\end{array}%
\right)\sim \left(1, 3, 2\right).
\end{equation}

Some flavons $\sigma^{u,d}$, $\chi^{e}$ and $\sigma^{\nu}$  are needed in the model and these are singlets under the gauge group but they transform in a non trivial way under the flavor symmetry. As can be noticed, the family indices have been neglected for simplicity.

Having introduced the matter content, in the lepton sector the type II see-saw mechanism will be used to get tiny neutrino masses, then, the Yukawa mass term is given by
\begin{equation}
\mathcal{L}=\mathcal{L}_{SM}-\frac{1}{2}Y^{\nu}\bar{L}(i\sigma_{2})\Delta \left(L\right)^{c}-V(H,\Delta)+h.c.
\end{equation}

In the present work, we ought to mention that the scalar potential analysis will be left out that we just focus in the Yukawa mass term. 

On the other hand, we would like to add some comments on the flavons and the flavor symmetry. For the first issue, the main role that the flavons will play is to generate non-renormalizable terms, in the quark as well as lepton sectors, which are crucial to get desirable mass textures. Speaking about the flavor symmetry, we will use the $\mathbf{S}_{3}$ due to the
three dimensional real representation can be decomposed as: ${\bf 3}_{S}={\bf 2}\oplus {\bf 1}_{S}$ or ${\bf 3}_{A}={\bf 2}\oplus {\bf 1}_{A}$. This structure seems to work quite well in the quark sector for obtaining hierarchical mass matrices. Along with this, the discrete symmetry, $\mathbf{Z}_{2}$, is required to prohibit some couplings in the quarks and lepton sectors. In order to keep this model minimal in its matter content and discrete symmetries, we have to make strong assumptions for each sector. This is a weakness about our model but many interesting features will be pointed out in the conclusions.

\subsection{QUARK SECTOR}
As it is well known, the CKM mixing matrix is fitted quite well by the NNI mass textures. Remarkably, these kind of textures may be generated by the $\mathbf{S}_{3}$ flavor symmetry where extra Higgs doublets are needed~\cite{Canales:2013cga}\footnote{In the previous cite, the authors studied the quark sector with three Higgs doublets under the $\mathbf{S}_{3}$ symmetry. They obtained a Hermitian quark mass matrix and achieved to get the NNI textures by making one rotation on the quarks fields.}

In our model, we will consider three Higgs doublets and two flavon scalars that transform in a non trivial way under the flavor symmetry. To get the NNI mass textures,
we assign the first and second families of quarks to a flavor doublet and the third family to the flavor singlet. To prohibit some couplings, an extra discrete symmetry ($\mathbf{Z}_{2}$) has been added. Thus, the assignation for the matter content is displayed in the following table.

\begin{table}[ht]
\begin{center}
\begin{tabular}{|c|c|c|c|c|c|c|c|c|c|c|}
\hline\hline
{\footnotesize Matter}	& {\footnotesize $Q_{I L}$} & {\footnotesize $Q_{3 L}$} & {\footnotesize $d_{I R}$}  & {\footnotesize $d_{3 R}$} & {\footnotesize $u_{I R}$} & {\footnotesize $u_{3 R}$} & {\footnotesize $H_{I}$}  & {\footnotesize $H_{3}$} & {\footnotesize $\sigma^{d}$} &  {\footnotesize $\sigma^{u}$} \\
\hline
{\footnotesize $\mathbf{S}_{3}$}	& {\footnotesize \bf $2$} & {\footnotesize \bf $1_{S}$} & {\footnotesize \bf $2$} & {\footnotesize \bf $1_{S}$} & {\footnotesize \bf $2$} & {\footnotesize \bf $1_{S}$} & {\footnotesize \bf $2$} & {\footnotesize \bf $1_{S}$} & {\footnotesize \bf $1_{A}$} & {\footnotesize \bf $1_{A}$} \\
\hline
{\footnotesize  $\mathbf{Z}_{2}$}	& 1 & -1 & 1 & -1 & 1 & -1 & -1 & 1 & 1 & 1 \\\hline\hline
\end{tabular}\caption{Matter content for the quark sector}
	\end{center}	
\end{table}

The Yukawa couplings between the quark and Higgs fields, that are allowed by the gauge and flavor symmetry, are given by

\begin{align}
-\mathcal{L}_{Y}&=y^{d}\left[\bar{Q}_{1 L}H_{3}d_{1 R}+\bar{Q}_{2 L}H_{3} d_{2 R}\right]+y^{d}_{2}\left[\bar{Q}_{1 L}H_{1}+\bar{Q}_{2 L}H_{2}\right]d_{3 R}+y^{d}_{3}\bar{Q}_{3 L}\left[H_{1}d_{1 R}+H_{2}d_{2 R}\right]+y^{d}_{4}\bar{Q}_{3 L}H_{3}d_{3 R}\nn\\&+y^{u}_{i}
\left(H\rightarrow \tilde{H},d_{R}\rightarrow u_{R}\right)
+h.c.\label{eq2}
\end{align}

Notice the above Yukawa mass term provides a kind of mass matrices that are not compatible with the NNI textures. Then, at this stage, the flavon scalars take relevance so that we have to include a non-renormalizable term (one for the up and down quarks) as following

\begin{equation}
\mathcal{L}_{Y}\supset y^{d}_{1}\left[\bar{Q}_{1 L}H_{3}d_{2 R}-\bar{Q}_{2 L}H_{3} d_{1 R}\right]\frac{\sigma^{d}}{\Lambda}+ y^{u}_{1}\left[\bar{Q}_{1 L}\tilde{H}_{3}u_{2 R}-\bar{Q}_{2 L}\tilde{H}_{3} u_{1 R}\right]\frac{\sigma^{u}}{\Lambda}+h.c.
\label{12e}
\end{equation}

In the above expression, $\Lambda$ is the cut-off scale of the theory and we consider only terms at leading order in the cut-off scale. Having included the above terms, the quark mass matrix has the following structure

\begin{equation}
{\bf M}_{q}=\begin{pmatrix}
a^{\prime}_{q} & a_{q} & b_{q} \\ 
-a_{q} & a^{\prime}_{q} & b^{\prime}_{q} \\ 
c_{q} & c^{\prime}_{q} & f_{q}
\end{pmatrix},\label{eq1} 
\end{equation}

where the matrix elements are
\begin{eqnarray}
a_{q}&=&y^{q}_{1} \frac{ \langle\sigma^{q}\rangle}{\Lambda}\langle H_{3}\rangle,\quad b^{\prime}_{q}=y^{q}_{2} \langle H_{2}\rangle, \quad b_{q}=y^{q}_{2} \langle H_{1}\rangle;\nn\\  c_{q}&=&y^{q}_{3} \langle H_{1}\rangle,\quad c^{\prime}_{q}=y^{q}_{3} \langle H_{2}\rangle,\quad f_{q}=y^{q}_{4} \langle H_{3}\rangle,\quad a^{\prime}_{q}=y^{q} \langle H_{3}\rangle. \label{eq9}
\end{eqnarray}
Here, $q=u, d$.

It is time to remark the strong assumptions:
\begin{itemize}
\item There is no symmetry that prohibits to couple the flavor  $\sigma^{u}$ (or $\sigma^{d}$) to the down (up) sector.
	
\item There are two flavons $\sigma^{d, u}$ which have been assigned as $\mathbf{1}_{A}$ singlets under the $\mathbf{S}_{3}$ symmetry. Then, there are three non-renormalizable terms (for the up and down quarks) which are invariant under the flavor symmetry: $\bar{Q}_{I L}H_{3}d_{IR}(\sigma^{d}/\Lambda)$, $\bar{Q}_{I L}H_{I}d_{3}(\sigma^{d}/\Lambda)$ and $\bar{Q}_{3 L}H_{I}d_{I R}(\sigma^{d}/\Lambda)$ \footnote{If the last two terms were included, one would have the following mass matrix
\begin{equation}
{\bf \mathcal{M}}_{q}=\begin{pmatrix}
	a^{\prime}_{q} & a_{q} & b_{q}+k_{q} \\ 
	-a_{q} & a^{\prime}_{q} & b^{\prime}_{q}-k^{\prime}_{q} \\ 
	c_{q}+l_{q} & c^{\prime}_{q}-l^{\prime}_{q} & f_{q}
\end{pmatrix}.
\end{equation}
This matrix, as one can verify in a straightforward way, will have the same mass textures that ${\bf M}_{q}$ given in Eq. (\ref{eq1}) with the same assumptions.
} where $I=1, 2$ under the flavor symmetry. As can be noticed, we just included the former term to generate the entry $12$ in the quark mass matrices (see Eq. (\ref{12e})).

\item In the rest of the paper, we will put $a^{\prime}_{q}=0$ by hand and the main reason is to get the NNI textures.
\end{itemize}

\subsection{LEPTON SECTOR}

In order to maintain the minimal matter content, the type II see-saw mechanism is invoked to obtain tiny masses. Then, one triplet scalar will be included in the matter content. However, extra flavons are needed to generate the mass textures that provide the mixings. Analogously to the quark sector, under the flavor symmetry, we use the same assignation. 

Therefore, the assignation is shown in the following table.

\begin{table}[ht]
	\begin{center}
		\begin{tabular}{|c|c|c|c|c|c|c|c|c|c|c|}
			\hline\hline
			{\footnotesize Matter}	& {\footnotesize $L_{I}$} & {\footnotesize $L_{3}$} & {\footnotesize $e_{I R}$}  & {\footnotesize $e_{3 R}$} & {\footnotesize $H_{I}$} & {\footnotesize $H_{3}$} & {\footnotesize $\Delta$}  & {\footnotesize $\chi^{e}$} & {\footnotesize $\sigma^{\nu}_{I}$} \\
			\hline
			{\footnotesize $\mathbf{S}_{3}$}	& {\footnotesize \bf $2$} & {\footnotesize \bf $1_{S}$} & {\footnotesize \bf $2$} & {\footnotesize \bf $1_{S}$} & {\footnotesize \bf $2$} & {\footnotesize \bf $1_{S}$} & {\footnotesize \bf $1_{S}$} & {\footnotesize \bf $1_{S}$} & {\footnotesize \bf $2$}  \\
			\hline
			{\footnotesize  $\mathbf{Z}_{2}$}	& 1 & 1 & 1 & 1 & -1 & 1 & 1 & -1 & 1\\\hline\hline
		\end{tabular}\caption{Matter content for the lepton sector}
	\end{center}	
\end{table}

Given the matter content and its respective assignation, at leading order in the cut-off scale, the allowed Yukawa mass term is given as
\begin{eqnarray}
-\mathcal{L}_{Y}&=&y^{e}_{1}\left[\bar{L}_{1}\left(H_{1}e_{2 R}+H_{2}e_{1 R}\right)+\bar{L}_{2}\left(H_{1}e_{1 R}-H_{2}e_{2 R}\right)\right]\frac{\chi^{e}}{\Lambda}+y^{e}_{2}\left[\bar{L}_{1}H_{3}e_{1 R}+\bar{L}_{2}H_{3}e_{2 R}\right]+y^{e}_{3}\bar{L}_{3}H_{3}e_{3 R}\nn\\
&&+y^{\nu}_{1}\left[\bar{L}_{1}\left(\Delta\frac{\sigma^{\nu}_{1}}{\Lambda}L^{c}_{2}+ \Delta\frac{\sigma^{\nu}_{2}}{\Lambda}L^{c}_{1 }\right)+\bar{L}_{2}\left(\Delta\frac{\sigma^{\nu}_{1}}{\Lambda}L^{c}_{1}- \Delta\frac{\sigma^{\nu}_{2}}{\Lambda}L^{c}_{2}\right)\right]+y^{\nu}_{2}\left[\bar{L}_{1}\Delta L^{c}_{1}+\bar{L}_{2}\Delta L^{c}_{2}\right]\nn\\&&+
y^{\nu}_{3}\left[\bar{L}_{1}\Delta\frac{\sigma^{\nu}_{1}}{\Lambda}+ \bar{L}_{2}\Delta\frac{\sigma^{\nu}_{2}}{\Lambda}
\right]L^{c}_{3}+y^{\nu}_{4}\bar{L}_{3}\left[\Delta\frac{\sigma^{\nu}_{1}}{\Lambda}L^{c}_{1}+ \bar{L}_{2}\Delta\frac{\sigma^{\nu}_{2}}{\Lambda}L^{c}_{2}
\right]+y^{\nu}_{5}\bar{L}_{3}\Delta L^{c}_{3}+h.c.
\end{eqnarray}
As result of this, the charged lepton and neutrino mass matrices are given as

\begin{equation}
{\bf M}_{e}=\begin{pmatrix}
a_{e}+b^{\prime}_{e} & b_{e} & 0 \\ 
b_{e} & a_{e}-b^{\prime}_{e} & 0 \\ 
0 & 0 & c_{e}
\end{pmatrix},\qquad {\bf M}_{\nu}=\begin{pmatrix}
a_{\nu}+b^{\prime}_{\nu} & b_{\nu} & c_{\nu} \\ 
b_{\nu} & a_{\nu}-b^{\prime}_{\nu} & c^{\prime}_{\nu} \\ 
c_{\nu} & c^{\prime}_{\nu} & d_{\nu}
\end{pmatrix} 
\end{equation}
with the respective matrix elements given by
\begin{eqnarray}
a_{e}&=&y^{e}_{2}\langle H_{3}\rangle,\quad b^{\prime}_{e}=y^{e}_{1}\frac{\langle \chi^{e}\rangle}{\Lambda}\langle H_{2}\rangle,\quad b_{e}=y^{e}_{1}\frac{\langle \chi^{e}\rangle}{\Lambda}\langle H_{1}\rangle, \quad c_{e}=y^{e}_{3}\langle H_{3}\rangle,\quad a_{\nu}=y^{\nu}_{2}\langle \Delta \rangle, \nn\\ b^{\prime}_{\nu}&=&y^{\nu}_{1}\frac{\langle \sigma^{\nu}_{2}\rangle}{\Lambda}\langle \Delta\rangle,\quad
b_{\nu}=y^{\nu}_{1}\frac{\langle \sigma^{\nu}_{1}\rangle}{\Lambda}\langle \Delta\rangle,\quad c^{\prime}_{\nu}=y^{\nu}_{1}\frac{\langle \sigma^{\nu}_{2}\rangle}{\Lambda}\langle \Delta\rangle, \quad c_{\nu}=y^{\nu}_{1}\frac{\langle \sigma^{\nu}_{1}\rangle}{\Lambda}\langle \Delta\rangle,\quad d_{\nu}=y^{\nu}_{5}\langle \Delta \rangle.
\end{eqnarray}

In this sector, we have to keep in mind the following assumptions:
\begin{itemize}
\item For convenience, we will make the alignment $\langle \sigma^{\nu}_{I}\rangle=(\langle \sigma^{\nu}_{1} \rangle, 0 )$ in the vev's of the scalar fields and the main reason is to get  desired textures in the mass matrices.

\item The ratio $\langle \sigma^{\nu}_{1} \rangle/\Lambda\sim \lambda=0.225$ where the Wolfenstein parameter ($\lambda$) is a typical factor to suppress the matrix elements in non-renormalizeble models.
\end{itemize}

\section{Mixing Matrices}

\subsection{CKM MIXING MATRIX}

In this section we show you how to obtain the NNI textures. First at all, we have to keep in mind that a strong assumption has been made, this is, $a^{\prime}_{q}=0$. As a consequence, the number of the free parameters is reduced substantially to six ones. In fact, considering $\langle H_{1}\rangle= \langle H_{2}\rangle$, the quark mass matrix contains four free parameters since that $b^{\prime}_{q}=b_{q}$ and $c^{\prime}_{q}=c_{q}$.

Having aligned the vev's, the quark mass matrix is diagonalized by two bi-unitary matrices such that
${\bf U}^{\dagger}_{q L} {\bf M}_{q}
{\bf U}_{q R}=\hat{\bf M}_{q}$ where $\mathbf{\hat{M}}_{q}=\textrm{Diag.}(m_{q_{1}}, m_{q_{2}}, m_{q_{3}})$ with $m_{q_{i}}$ being the physical quark masses. Then, we make the following rotation ${\bf U}_{q(L, R)}={\bf U}_{\pi/4} {\bf u}_{q (L, R)}$
so that one obtains  $\mathbf{u}^{\dagger}_{q L} \mathbf{m}_{q}\mathbf{u}_{q R}=\mathbf{\hat{M}}_{q}$ where $\mathbf{m}_{q}$ and $\mathbf{U}_{\pi/4}$ are given respectively as

\begin{equation}
{\bf m}_{q}=
\begin{pmatrix}
0 & a_{q} & 0 \\ 
-a_{q} & 0 & B_{q} \\ 
0 & C_{q} & f_{q}
\end{pmatrix},
\qquad\mathbf{U}_{\pi/4}= \begin{pmatrix}
\frac{1}{\sqrt{2}} & \frac{1}{\sqrt{2}} & 0 \\ 
-\frac{1}{\sqrt{2}} & \frac{1}{\sqrt{2}} & 0 \\ 
0 & 0 & 1
\end{pmatrix}\label{eq4} 
\end{equation}

In this benchmark three free parameters can be fixed in terms ($\vert a_{q}\vert$, $\vert B_{q}\vert=\vert \sqrt{2}b_{q}\vert$  and $\vert C_{q}\vert=\vert \sqrt{2}c_{q}\vert$) of the physical masses and one unfixed free parameter, $y_{q}=\vert f_{q}\vert/m_{q_{3}}$. Therefore, we have to figure out the form of the ${\bf u}_{f R}$ and ${\bf u}_{f L}$
unitary matrices that diagonalize ${\bf m}_{q}$. Then, we must build the
bilineal forms: ${\bf \hat{M}}_{q} {\bf \hat{M}}^{\dagger}_{q}={\bf
	u}^{\dagger}_{q L} {\bf m}_{q}{\bf m }^{\dagger}_{q} {\bf u}_{q L}$
and ${\bf \hat{M}}^{\dagger}_{q} {\bf \hat{M}}_{q}={\bf
	u}^{\dagger}_{q R}{\bf m}^{\dagger}_{q}{\bf m }_{q} {\bf u}_{q
	R}$, however, in this work we will only need to  obtain the
${\bf u}_{q L}$ left-handed matrix which takes place in the CKM
matrix. This is given by ${\bf u}_{q L}={\bf Q}_{q L}{\bf O}_{q L}$
where the former matrix contains the CP-violating phases, $ {\bf
	Q}_{q} = \textrm{diag} \left( 1,\exp i\bar{\alpha}_{q}, \exp
i\bar{\beta}_{q} \right)$, that comes from ${\bf m}_{q}{\bf m
}^{\dagger}_{q}$. ${\bf O}_{q L}$ is a real orthogonal matrix and this
is parametrized as
\begin{align} \label{ortho}
{\bf O}_{q L}= 
\begin{pmatrix}
-\sqrt{\dfrac{\tilde{m}_{q_{2}} (\rho^{q}_{-}-R^{q}) K^{q}_{+}}{4 y_{q} \delta^{q}_{1} \kappa^{q}_{1} }} & -\sqrt{\dfrac{\tilde{m}_{q_{1}} (\sigma^{q}_{+}-R^{q}) K^{f}_{+}}{4 y_{q} \delta^{q}_{2} \kappa^{q}_{2} }}  & \sqrt{\dfrac{\tilde{m}_{q_{1}} \tilde{m}_{q_{2}} (\sigma^{q}_{-}+R^{q}) K^{q}_{+}}{4 y_{q} \delta^{q}_{3} \kappa^{q}_{3} }} \\ 
-\sqrt{\dfrac{\tilde{m}_{q_{1}} \kappa^{q}_{1} K^{q}_{-}}{\delta^{q}_{1}(\rho^{q}_{-}-R^{q}) }} & \sqrt{\dfrac{\tilde{m}_{q_{2}} \kappa^{q}_{2} K^{q}_{-}}{\delta^{q}_{2}(\sigma^{q}_{+}-R^{q}) }} & \sqrt{\dfrac{\kappa^{q}_{3} K^{q}_{-}}{\delta^{q}_{3}(\sigma^{q}_{-}+R^{q}) }} \\ 
\sqrt{\dfrac{\tilde{m}_{q_{1}} \kappa^{q}_{1}(\rho^{q}_{-}-R^{q})}{2 y_{q}\delta^{q}_{1}}} & -\sqrt{\dfrac{\tilde{m}_{q_{2}} \kappa^{q}_{2}(\sigma^{q}_{+}-R^{q})}{2 y_{q}\delta^{q}_{2}}}  & \sqrt{\dfrac{\kappa^{q}_{3}(\sigma^{q}_{-}+R^{q})}{2 y_{q}\delta^{q}_{3}}}
\end{pmatrix}
\end{align}
with
\begin{align} \label{defortho}
\rho^{q}_{\pm}&\equiv 1+\tilde{m}^{2}_{q_{2}}\pm\tilde{m}^{2}_{q_{1}}-y^{2}_{q},\quad \sigma^{q}_{\pm}\equiv 1-\tilde{m}^{2}_{q_{2}}\pm(\tilde{m}^{2}_{q_{1}}-y^{2}_{q}),\quad
\delta^{q}_{(1, 2)}\equiv (1-\tilde{m}^{2}_{q_{(1, 2)}})(\tilde{m}^{2}_{q_{2}}-\tilde{m}^{2}_{q_{1}});\nn\\
\delta^{q}_{3}&\equiv (1-\tilde{m}^{2}_{q_{1}})(1-\tilde{m}^{2}_{q_{2}}),\quad \kappa^{q}_{1} \equiv  \tilde{m}_{q_{2}}-\tilde{m}_{q_{1}}y_{q},\quad \kappa^{q}_{2}\equiv \tilde{m}_{q_{2}}y_{q}-\tilde{m}_{q_{1}},\quad \kappa^{q}_{3}\equiv y_{q}-\tilde{m}_{q_{1}}\tilde{m}_{q_{2}};\nn\\
R^{q}&\equiv \sqrt{\rho^{q 2}_{+}-4(\tilde{m}^{2}_{q_{2}}+\tilde{m}^{2}_{q_{1}}+\tilde{m}^{2}_{q_{2}}\tilde{m}^{2}_{q_{1}}-2\tilde{m}_{q_{1}}\tilde{m}_{q_{2}}y_{q})},\quad	K^{q}_{\pm} \equiv  y_{q}(\rho^{q}_{+}\pm R^{q})-2\tilde{m}_{q_{1}}\tilde{m}_{q_{2}}.
\end{align}

In the above expressions, all the parameters have been normalized by
the heaviest physical quark mass, $m_{q_{3}}$. Along with this, from
the above parametrization, $y_{q}\equiv \vt f_{q}\vt/m_{q_{3}}$, is
the only dimensionless free parameter that cannot be fixed in terms of
the physical masses, but is is constrained by, 
$1>y_{q}>\tilde{m}_{q_{2}}>\tilde{m}_{q_{1}}$. Therefore, the
left-handed mixing matrix that takes places in the CKM matrix is given
by ${\bf U}_{q L}= {\bf U}_{\pi/4}{\bf Q}_{q}{\bf O}_{q L}$ where
$q=u, d$. Finally, the CKM mixing matrix is written as
\begin{equation}
{\bf V}_{CKM}={\bf O}^{T}_{u L}{\bf P}_{q} {\bf O}_{d L}, \quad {\bf P}_{q}={\bf Q}^{\dagger}_{u}{\bf Q}_{d}=\textrm{diag.}\left(1, e^{i\alpha_{q}}, e^{i\beta_{q}} \right).
\end{equation}

As one can notice, the CKM mixing matrix has four free parameters, namely $y_{u}$, $y_{d}$, and two phases $\alpha_{q}$ and $\beta_{q}$. These parameters will be constrained only numerically since that an analytical study on the NNI textures was carried out in the past ~\cite{Branco:1988iq, Branco:1994jx, Harayama:1996am, Harayama:1996jr} to find a set of values for the free parameters that fit the CKM mixing matrix.

\subsection{PMNS MIXING MATRIX}
As a result of choosing the alignment $\langle H_{1}\rangle= \langle H_{2}\rangle$ in the vev's, then in the charged lepton mass matrix one gets $b_{e}=b^{\prime}_{e}$. Although this alignment allows us to reduce the free parameters in the quark and charged lepton sector, in the neutrino sector there are too many. Hence, let us consider the following vev's for the doublets $\langle \sigma^{\nu}_{I}\rangle=(\langle \sigma^{\nu}_{1} \rangle, 0 )$. With this in mind, the mass matrices are give as

\begin{equation}
{\bf M}_{e}=\begin{pmatrix}
a_{e}+b_{e} & b_{e} & 0 \\ 
b_{e} & a_{e}-b_{e} & 0 \\ 
0 & 0 & c_{e}
\end{pmatrix},\qquad {\bf M}_{\nu}=\begin{pmatrix}
a_{\nu} & b_{\nu} & c_{\nu} \\ 
b_{\nu} & a_{\nu} & 0 \\ 
c_{\nu} & 0 & d_{\nu}
\end{pmatrix} 
\end{equation}

Due to the above alignment, the neutrino mass matrix possesses a kind of Fritzsch textures but there is an extra parameter as we will see next. The charged lepton and neutrino mass matrix are diagonalized by ${\bf U}^{\dagger}_{e L} {\bf M}_{e}
{\bf U}_{e R}=\hat{\bf M}_{e}$ where $\mathbf{\hat{M}}_{e}=\textrm{Diag.}(m_{e}, m_{\mu}, m_{\tau})$. In the neutrino sector, we have ${\bf U}^{\dagger}_{\nu} {\bf M}_{\nu}
{\bf U}^{\ast}_{\nu}=\hat{\bf M}_{\nu}$ with $\mathbf{\hat{M}}_{\nu}=\textrm{Diag.}(m_{1}, m_{2}, m_{3})$ being the physical neutrino masses. Then, for both sectors, we make the following rotation ${\bf U}_{e(L, R)}={\bf S}_{12} {\bf u}_{e (L, R)}$ and ${\bf U}_{\nu}={\bf S}_{12} {\bf u}_{\nu}$.
so that one obtains  $\mathbf{u}^{\dagger}_{e L} \mathbf{m}_{e}\mathbf{u}_{e R}=\mathbf{\hat{M}}_{e}$  and $\mathbf{u}^{\dagger}_{\nu} \mathbf{m}_{\nu}\mathbf{u}^{\ast}_{\nu }=\mathbf{\hat{M}}_{\nu}$  where $\mathbf{m}_{e}$, $\mathbf{m}_{\nu}$ and $\mathbf{S}_{12}$ are given respectively as

\begin{equation}
{\bf m}_{e}=\begin{pmatrix}
a_{e}-b_{e} & b_{e} & 0 \\ 
b_{e} & a_{e}+b_{e} & 0 \\ 
0 & 0 & c_{e}
\end{pmatrix},\qquad {\bf m}_{\nu}=\begin{pmatrix}
a_{\nu} & b_{\nu} & 0 \\ 
b_{\nu} & a_{\nu} & c_{\nu} \\ 
0 & c_{\nu} & d_{\nu}
\end{pmatrix},\qquad \mathbf{S}_{12}= \begin{pmatrix}
0 & 1 & 0 \\ 
1 & 0 & 0 \\ 
0 & 0 & 1
\end{pmatrix}
\end{equation}

As one can realize, if $a_{\nu}$ was zero, the mass matrix, $\mathbf{m}_{\nu}$, would possess the Fritzsch textures. Then, this extra parameter, $a_{q}$, will modify slightly the Fritzsch texture as we will see in the diagonalization procedure. 

As one can verify straightforward, the charged lepton mass matrix is diagonalized by $\mathbf{u}_{e L}=\mathbf{P}_{e}\mathbf{O}_{e}$ and $\mathbf{u}_{e R}=\mathbf{P}^{\dagger}_{e}\mathbf{O}_{e}$ where $\mathbf{P}_{e}$ and $\mathbf{O}_{e}$ are a diagonal phase and orthogonal real mass matrices, respectively. Explicitly, we get
\begin{eqnarray}
\mathbf{P}_{e}=\begin{pmatrix}
e^{i\eta_{e}}& 0 & 0 \\
0 & e^{i\eta_{\mu}} & 0 \\
0 & 0 & e^{i\eta_{\tau}}
\end{pmatrix},\qquad \mathbf{O}_{e}=\begin{pmatrix}
\cos{\theta_{e}} & \sin{\theta_{e}} & 0 \\
-\sin{\theta_{e}} & \cos{\theta_{e}} & 0 \\
0 & 0 & 1
\end{pmatrix}
\end{eqnarray}
with
\begin{eqnarray}
\eta_{e}&=&\frac{arg(a_{e}-b_{e})}{2}, \qquad \eta_{\mu}=\frac{arg(a_{e}+b_{e})}{2},\quad \eta_{\tau}=\frac{arg(c_{e})}{2}\qquad \textrm{and}\qquad \eta_{e}+\eta_{\mu}=arg(b_{e});\nn\\
\cos{\theta_{e}}&=&\sqrt{\frac{m_{\mu}-m_{e}+R_{e}}{2(m_{\mu}-m_{e})}},\qquad \sin{\theta_{e}}=\sqrt{\frac{m_{\mu}-m_{e}-R_{e}}{2(m_{\mu}-m_{e})}},\qquad R_{e}=\sqrt{(m_{\mu}-m_{e})^{2}-4\vert b_{e}\vert^{2}}.
\end{eqnarray}
Here, $\vert b_{e}\vert$ is a free parameter which is constrained by the following relation $m_{\mu}>\vert b_{e}\vert>m_{e}$.

Now, let us focus on the neutrino mass matrix, $\mathbf{m}_{\nu}$. First at all, the CP violating phases are factorized as $\mathbf{m}_{\nu}=\mathbf{P}_{\nu}\bar{\mathbf{m}}_{\nu}\mathbf{P_{\nu}}$. Explicitly, we obtain

\begin{eqnarray}
\mathbf{P}_{\nu}=\begin{pmatrix}
e^{i\eta_{1}}& 0 & 0 \\
0 & e^{i\eta_{2}} & 0 \\
0 & 0 & e^{i\eta_{3}}
\end{pmatrix},\qquad \bar{\mathbf{m}}_{\nu}=\begin{pmatrix}
\vert a_{\nu}\vert & \vert b_{\nu}\vert & 0 \\
\vert b_{\nu}\vert & \vert a_{\nu}\vert & \vert c_{\nu}\vert \\
0 & \vert c_{\nu}\vert  & \vert d_{\nu}\vert 
\end{pmatrix}
\end{eqnarray}
with the following condition on the CP phases
\begin{equation}
\eta_{1}=\frac{arg(a_{\nu})}{2},\quad \eta_{2}=\frac{arg(a_{\nu})}{2},\quad \eta_{3}=\frac{arg(d_{\nu})}{2},\quad \eta_{1}+\eta_{2}=arg(b_{\nu}),\quad \eta_{2}+\eta_{3}=arg(c_{\nu}).
\end{equation} 

As a result of factorizing the CP violating phases, we have that $\mathbf{u}_{\nu}=\mathbf{P}_{\nu}\mathbf{O}_{\nu}$. Let us obtain the orthogonal matrix that diagonalizes the real symmetric mass matrix, $\bar{\mathbf{m}}_{\nu}$. Here, we will consider two cases: the normal and inverted hierarchy in the neutrino masses.

\subsubsection{Normal Hierarchy (NH)}

Given the neutrino mass matrix, we can fix three free parameters in terms of the neutrino physical masses and one unfixed free parameter, $\vert a_{\nu}\vert$. This is,
\begin{eqnarray}
\vert d_{\nu}\vert&=& m_{3}-\vert m_{2}\vert+m_{1}-2\vert a_{\nu}\vert\nn\\
\vert b_{\nu}\vert&=&\sqrt{\frac{(m_{3}-\vert a_{\nu}\vert)(\vert m_{2}\vert+\vert a_{\nu}\vert)(m_{1}-\vert a_{\nu}\vert)}{m_{3}-\vert m_{2}\vert+m_{1}-3\vert a_{\nu}\vert}}\nn\\
\vert c_{\nu}\vert&=&\sqrt{\frac{(m_{3}+m_{1}-2\vert a_{\nu}\vert)(m_{3}-\vert m_{2}\vert-2\vert a_{\nu}\vert)(\vert m_{2}\vert-m_{1}+2\vert a_{\nu}\vert)}{m_{3}-\vert m_{2}\vert +m_{1}-3\vert a_{\nu}\vert}}
\end{eqnarray}
Here, we have taken $m_{2}=-\vert m_{2}\vert$ in order to get real parameters. Along with this, there is a constraint for the unfixed free parameter $m_{3}>\vert m_{2}\vert>m_{1}>\vert a_{\nu}\vert>0$. After a lengthily task, we obtain the orthogonal real matrix
\begin{equation}
\mathbf{O}_{\nu}=\begin{pmatrix}
\sqrt{\frac{( \tilde{m}_{2}+\tilde{a}_{\nu})(1-\tilde{a}_{\nu})\mathcal{M}_{2}}{\mathcal{D}_{1}}}& -\sqrt{\frac{(\tilde{m}_{{1}}-\tilde{a}_{\nu})(1-\tilde{a}_{\nu})\mathcal{M}_{1}}{\mathcal{D}_{2}}}
& \sqrt{\frac{(\tilde{m}_{2}+\tilde{a}_{\nu})(\tilde{m}_{1}-\tilde{a}_{\nu})\mathcal{M}_{3}}{\mathcal{D}_{3}}} 
\\ 
\sqrt{\frac{(\tilde{m}_{1}-\tilde{a}_{\nu})\mathcal{M}_{2}\mathcal{D}}{\mathcal{D}_{1}}}& \sqrt{\frac{(\tilde{m}_{2}+\tilde{a}_{\nu})\mathcal{M}_{1}\mathcal{D}}{\mathcal{D}_{2}}}
& \sqrt{\frac{(1-\tilde{a}_{\nu})\mathcal{M}_{3}\mathcal{D}}{\mathcal{D}_{3}}} \\ 
-\sqrt{\frac{(\tilde{m}_{1}-\tilde{a}_{\nu})\mathcal{M}_{1}\mathcal{M}_{3}}{\mathcal{D}_{1}}}&-\sqrt{\frac{(\tilde{m}_{2}+\tilde{a}_{\nu})\mathcal{M}_{2}\mathcal{M}_{3}}{\mathcal{D}_{2}}} 
& 
\sqrt{\frac{(1-\tilde{a}_{\nu})\mathcal{M}_{1}\mathcal{M}_{2}}{\mathcal{D}_{3}}}
\label{eq7}
\end{pmatrix} 
\end{equation}
with 
\begin{eqnarray}
\mathcal{M}_{1}&=&1+\tilde{m}_{1}-2\tilde{a}_{\nu},\quad 
\mathcal{M}_{2}=1-\tilde{m}_{2}-2\tilde{a}_{\nu},\quad
\mathcal{M}_{3}=\tilde{m}_{2}-\tilde{m}_{1}+2\tilde{a}_{\nu}
,\quad
\mathcal{D}=1-\tilde{m}_{2}+\tilde{m}_{1}-3\tilde{a}_{\nu};\nonumber\\
\mathcal{D}_{1}&=&(1-\tilde{m}_{1})( \tilde{m}_{2}+\tilde{m}_{1})\mathcal{D},\quad
\mathcal{D}_{2}=(1+\tilde{m}_{2})( \tilde{m}_{2}+\tilde{m}_{1})\mathcal{D},\quad
\mathcal{D}_{3}=(1+ \tilde{m}_{2})(1-\tilde{m}_{1})\mathcal{D},
\label{eq8}
\end{eqnarray}
where $\tilde{m}_{{2}}=\vert m_{{2}}\vert /m_{3}$, $\tilde{m}_{1}=m_{1}/m_{3}$ and $\tilde{a}_{\nu}=\vert a_{\nu}\vert/ m_{3}$. In this parametrization, there is a constraint among the quark masses and the free parameter $\tilde{a}_{\nu}$, this is $1>\tilde{m}_{2}>\tilde{m}_{1}>\tilde{a}_{\nu}>0$. As we observed, for simplicity, the mixing matrix elements in have been normalized by the heaviest mass.

\subsubsection{Inverted Hierarchy (IH)}

In this case, the fixed free parameters
\begin{eqnarray}
\vert d_{\nu}\vert&=& m_{2}-\vert m_{1}\vert+m_{3}-2\vert a_{\nu}\vert\nn\\
\vert b_{\nu}\vert&=&\sqrt{\frac{(m_{3}-\vert a_{\nu}\vert)(\vert m_{1}\vert+\vert a_{\nu}\vert)(m_{2}-\vert a_{\nu}\vert)}{m_{2}-\vert m_{1}\vert+m_{3}-3\vert a_{\nu}\vert}}\nn\\
\vert c_{\nu}\vert&=&\sqrt{\frac{(\vert m_{1}\vert-m_{3}+2\vert a_{\nu}\vert)(m_{2}+ m_{3}-2\vert a_{\nu}\vert)(m_{2}-\vert m_{1}\vert-2\vert a_{\nu}\vert)}{m_{2}-\vert m_{1}\vert +m_{3}-3\vert a_{\nu}\vert}}
\end{eqnarray}
For this ordering in the neutrino masses, we have taken $m_{1}=-\vert m_{1}\vert$ for getting the three parameters reals. Along with this, the orthogonal real matrix is given by
\begin{equation}
\mathbf{O}_{\nu}=\begin{pmatrix}
-\sqrt{\frac{(1-\tilde{a}_{\nu})(\tilde{m}_{3}-\tilde{a}_{\nu})\mathcal{N}_{2}}{\mathcal{D}_{\nu_{1}}}}& \sqrt{\frac{(\tilde{m}_{{1}}+\tilde{a}_{\nu})(\tilde{m}_{3}-\tilde{a}_{\nu})\mathcal{N}_{1}}{\mathcal{D}_{\nu_{2}}}}
& \sqrt{\frac{(1-\tilde{a}_{\nu})(\tilde{m}_{1}+\tilde{a}_{\nu})\mathcal{N}_{3}}{\mathcal{D}_{\nu_{3}}}} 
\\ 
\sqrt{\frac{(\tilde{m}_{1}+\tilde{a}_{\nu})\mathcal{N}_{2}\mathcal{D}_{\nu}}{\mathcal{D}_{\nu_{1}}}}& \sqrt{\frac{(1-\tilde{a}_{\nu})\mathcal{N}_{1}\mathcal{D}_{\nu}}{\mathcal{D}_{\nu_{2}}}}
& \sqrt{\frac{(\tilde{m}_{3}-\tilde{a}_{\nu})\mathcal{N}_{3}\mathcal{D}_{\nu}}{\mathcal{D}_{\nu_{3}}}} \\ 
-\sqrt{\frac{(\tilde{m}_{1}+\tilde{a}_{\nu})\mathcal{N}_{1}\mathcal{N}_{3}}{\mathcal{D}_{\nu_{1}}}}&\sqrt{\frac{(1-\tilde{a}_{\nu})\mathcal{N}_{2}\mathcal{N}_{3}}{\mathcal{D}_{\nu_{2}}}} 
& 
-\sqrt{\frac{(\tilde{m}_{3}-\tilde{a}_{\nu})\mathcal{N}_{1}\mathcal{N}_{2}}{\mathcal{D}_{\nu_{3}}}}
\label{eq88}
\end{pmatrix} 
\end{equation}
where
\begin{eqnarray}
\mathcal{N}_{1}&=&\tilde{m}_{1}-\tilde{m}_{3}+2\tilde{a}_{\nu},\quad 
\mathcal{N}_{2}=1+\tilde{m}_{3}-2\tilde{a}_{\nu},\quad
\mathcal{N}_{3}=1-\tilde{m}_{1}-2\tilde{a}_{\nu}
,\quad
\mathcal{D}_{\nu}=1-\tilde{m}_{1}+\tilde{m}_{3}-3\tilde{a}_{\nu};\nonumber\\
\mathcal{D}_{\nu_{1}}&=&(1+\tilde{m}_{1})( \tilde{m}_{1}+\tilde{m}_{3})\mathcal{D}_{\nu},\quad
\mathcal{D}_{\nu_{2}}=(1+\tilde{m}_{1})( 1-\tilde{m}_{3})\mathcal{D}_{\nu},\quad
\mathcal{D}_{\nu_{3}}=(1- \tilde{m}_{3})(\tilde{m}_{1}+\tilde{m}_{3})\mathcal{D}_{\nu},
\label{eq99}
\end{eqnarray}
where $\tilde{m}_{{1}}=\vert m_{{1}}\vert /m_{2}$, $\tilde{m}_{3}=m_{3}/m_{2}$ and $\tilde{a}_{\nu}=\vert a_{\nu}\vert/ m_{2}$. In this parametrization, there is a constraint among the quark masses and the free parameter $\tilde{a}_{\nu}$, this is $1>\tilde{m}_{1}>\tilde{m}_{3}>\tilde{a}_{\nu}>0$.

So, we end up having the PMNS mixing matrix $\mathbf{V}^{i}=\mathbf{U}^{\dagger}_{e L}\mathbf{U}^{i}_{\nu}=\mathbf{O}^{T}_{e}\bar{\mathbf{P}}_{e}\mathbf{O}^{i}_{\nu}$. In here, $i= NH, IH$, in addition, $\bar{\mathbf{P}}_{e}=\mathbf{P}^{\dagger}_{e}\mathbf{P}_{\nu}\equiv\textrm{Diag.}(e^{i\bar{\eta}_{e}}, e^{i\bar{\eta}_{\mu}},e^{i\bar{\eta}_{\tau}})$. Thus, we can compare our expression with the standard parametrization of the PMNS mixing matrix such that the reactor, atmospheric and solar angles are well determined by
\begin{eqnarray}\label{mixan}
\sin{\theta}_{13}&=&\vert (\mathbf{V}^{i})_{13}\vert=\vert \cos{\theta}_{e}(\mathbf{O}^{i}_{\nu})_{13}e^{i\bar{\eta}_{e}}-\sin{\theta}_{e}(\mathbf{O}^{i}_{\nu})_{23} e^{i\bar{\eta}_{\mu}}\vert\nn\\
\sin{\theta}_{23}&=&\frac{\vert (\mathbf{V}^{i})_{23}\vert}{\sqrt{1-\sin^{2}{\theta}_{13}}}=\frac{\vert \sin{\theta}_{e}(\mathbf{O}^{i}_{\nu})_{13}e^{i\bar{\eta}_{e}}+\cos{\theta_{e}}(\mathbf{O}^{i}_{\nu})_{23} e^{i\bar{\eta}_{\mu}}\vert}{\sqrt{1-\sin^{2}{\theta}_{13}}}\nn\\
\sin{\theta}_{12}&=&\frac{\vert (\mathbf{V}^{i})_{12}\vert}{\sqrt{1-\sin^{2}{\theta}_{13}}}=\frac{\vert \cos{\theta}_{e}(\mathbf{O}^{i}_{\nu})_{12} e^{i\bar{\eta}_{e}}-\sin{\theta_{e}}(\mathbf{O}^{i}_{\nu})_{22} e^{i\bar{\eta}_{\mu}}\vert}{\sqrt{1-\sin^{2}{\theta}_{13}}}
\end{eqnarray}

Notice that in the PMNS matrix there are five free parameters namely: $\vert b_{e}\vert$, $\vert a_{\nu}\vert$ and three CP violating phases but only one takes place effectively, $\eta_{\nu}\equiv \bar{\eta}_{\mu}-\bar{\eta}_{e}$, in Eq.(\ref{mixan}). In fact, due to of lacking information on the absolute neutrino masses, the lightest one may be considered as an extra free parameter.

On the other hand, we would like to point out a little comment on the Majorana phases. In this model, for the normal ordering, we have considered the particular case $(m_{3}, m_{2}, m_{1})=(+,-,+)$. In the inverted hierarchy, we have $(m_{3}, m_{2}, m_{1})=(+,+,-)$.

\section{Results}

\subsection{CKM MIXING MATRIX}
As we already commented, the quark mass matrices possess the NNI textures, as it is well known, these reproduce quite well the CKM mixing matrix so that we will not do a analytical study. Instead of doing that a numerical analysis will be carried out.

\begin{table}[th]
\begin{center}
\begin{tabular}{|c|c|c}
\hline\hline
Observable & Experimental value \\ \hline
$m_{u}(MeV)$ & \quad $1.45_{-0.45}^{+0.56}$ \\ \hline
$m_{c}(MeV)$ &  \quad $635\pm 86$ \\ \hline
$m_{t}(GeV)$ &  \quad $172.1\pm 0.6\pm 0.9$ \\ \hline
$m_{d}(MeV)$ & \quad $2.9_{-0.4}^{+0.5}$ \\ \hline
$m_{s}(MeV)$ & \quad $57.7_{-15.7}^{+16.8}$ \\ \hline
$m_{b}(GeV)$ & \quad $2.82_{-0.04}^{+0.09}$ \\ \hline
$V_{ud}$ & \quad $0.97434^{+0.00011}_{-0.00012}$ \\ \hline
$V_{us}$ &  \quad $0.22506\pm 0.00050 $ \\ \hline
$V_{ub}$ & \quad $0.00357\pm 0.00015 $ \\ \hline
$J_q$ &  $\left(3.18\pm 0.15\right)\times 10^{-5}$ \\ \hline\hline
\end{tabular}%
\end{center}
\caption{ Experimental values of the quark masses \cite{Bora:2012tx} and CKM parameters \cite{Patrignani:2016xqp}.}
	\label{Tab}
\end{table}

In this case, the CKM mixing matrix contains four free parameters namely
$y_{u}$, $y_{d}$ and two CP violating phases where the former ones satisfy the constraint $1>y_{q}>\tilde{m}_{q_{2}}>\tilde{m}_{q_{1}}$ for each sector; and the two phases are in the interval of $\left[0, 2\pi\right]$. In this work, the physical quark masses (at
$m_{Z}$ scale) will be taken (just central values) as inputs. Thus, in the following, a naive $\chi^{2}$ analysis will
be performed to tune the free parameters. Then, we define
\begin{equation}
\chi^{2}\left(y_{u},y_{d}, \eta_{q_{1}}, \eta_{q_{2}}\right)=\frac{\left(\left| V^{th}_{ud}\right|-V^{ex}_{ud}\right)^{2}}{\sigma^{2}_{ud}}+\frac{\left(\left|V^{th}_{us}\right|-V^{ex}_{us}\right)^{2}}{\sigma^{2}_{us}}+\frac{\left(\left|V^{th}_{ub}\right|-V^{ex}_{ub}\right)^{2}}{\sigma^{2}_{ub}}+\frac{\left(\left|J^{th}\right|-J^{ex}\right)^{2}}{\sigma^{2}_{J}}.
\end{equation}
where 
\begin{equation}
J^{th}=Im\left[V^{th}_{us}~V^{th}_{cb}~ V^{\ast th}_{cs}~ V^{\ast th}_{ub} \right].\nonumber
\end{equation}
Then, we obtain the following values for the free parameters that fit the mixing values up to $2\sigma$
\begin{equation}
y_{u}=0.996068,\quad y_{d}=0.922299,\quad \eta_{q_{1}}=4.48161,\quad \eta_{q_{2}}=3.64887,
\end{equation}
with these values, one obtains
\begin{equation}
\left|V^{th}_{CKM}\right|=\begin{pmatrix}
0.97433 & 0.22505 & 0.00356 \\ 
0.22490 & 0.97359 & 0.03926 \\ 
0.00901 & 0.03831 & 0.99922
\end{pmatrix},\qquad J^{th}=3.04008\times 10^{-5}.
\end{equation}
As can be seen, these values are in good agreement with the experimental data, this is not a surprise since the NNI textures work quite well in the quark sector.

\subsection{PMNS MIXING MATRIX}

\subsubsection{Analytically study}

In order to try of figuring out the allowed region for free parameters, let us make a brief analytical study on the mixing angles formulas. First at all, for the normal and inverted hierarchy, two neutrino masses can be fixed in terms of the squared mass scales and the lightest neutrino mass. This is,
\begin{eqnarray}\label{massfix}
m_{3}&=&\sqrt{\Delta m^{2}_{31}+m^{2}_{1}},\qquad \vert m_{2}\vert=\sqrt{\Delta m^{2
	}_{21}+m^{2}_{1}},\qquad \textrm{Normal Hierarchy}\nn\\
m_{2}&=&\sqrt{\Delta m^{2}_{23}+m^{2}_{3}},\qquad \vert m_{1}\vert =\sqrt{\Delta m^{2
	}_{23}-\Delta m^{2}_{21}+m^{2}_{3}}.\qquad \textrm{Inverted Hierarchy}
\end{eqnarray}

Then, for each case, the lightest neutrino mass is an extra free parameter that needs to be constrained.
Having commented this, in the PMNS mixing matrix there is a free parameter, $\tilde{a}_{\nu}$, which is constrained according to the mass spectrum. So that, we will study two extreme values for that parameter in order to have an idea of the mixing angles values.
\begin{description}
\item [Normal Hierarchy]
 ($1>\tilde{m}_{2}>\tilde{m}_{1}>\tilde{a}_{\nu}>0$).

As we can notice, the neutrino mixing matrix has been normalized by the heaviest neutrino mass. In this case,
from Eq. (\ref{massfix}), we have the normalize neutrino masses
\begin{equation}\label{mr}
\tilde{m}_{2}=\sqrt{\frac{\Delta m^{2
		}_{21}+m^{2}_{1}}{\Delta m^{2
		}_{31}+m^{2}_{1}}},\qquad \tilde{m}_{1}=\frac{m_{1}}{\sqrt{\Delta m^{2
		}_{31}+m^{2}_{1}}}, \qquad \frac{\tilde{m}_{1}}{\tilde{m}_{2}}=\frac{m_{1}}{\sqrt{\Delta m^{2
		}_{21}+m^{2}_{1}}}.
\end{equation}
The above expression will be useful to fix ideas about the magnitude of the mixing angles.

\begin{itemize}
	\item Case I: $\tilde{a}_{\nu}\approx 0$. In this case, the neutrino mass matrix corresponds to the Fritzsch textures and the involved entries of the mixing matrix are given as
	\begin{eqnarray}
	(\mathbf{O}_{\nu})_{13}&\approx& \tilde{m}_{2}\sqrt{\tilde{m}_{1}\left(1+\tilde{m}^{2}_{2}-\frac{\tilde{m}_{1}}{\tilde{m}_{2}}\right)}, \qquad
	(\mathbf{O}_{\nu})_{23}\approx\sqrt{\tilde{m}_{2}\left(1-\frac{\tilde{m}_{1}}{\tilde{m}_{2}}-\tilde{m}_{2}\right)};\nn\\
	(\mathbf{O}_{\nu})_{12}&\approx&- \sqrt{\frac{\tilde{m}_{1}}{\tilde{m}_{2}}\left(1+\tilde{m}^{2}_{2}-\frac{\tilde{m}_{1}}{\tilde{m}_{2}}\right)},\qquad
	(\mathbf{O}_{\nu})_{22}\approx\sqrt{1-\frac{\tilde{m}_{1}}{\tilde{m}_{2}}-\tilde{m}_{2}}.
	 \end{eqnarray}

Then, in magnitude, we have that $\vert (\mathbf{O}_{\nu})_{13}\vert\ll \vert(\mathbf{O}_{\nu})_{23}\vert$ and  $\vert (\mathbf{O}_{\nu})_{12}\vert\ll \vert(\mathbf{O}_{\nu})_{22}\vert$. As result of this, for the reactor and atmospheric angles, the main contribution comes from the entry $(\mathbf{O}_{\nu})_{23}$. Along with this, the solar angle depends strongly on the two entries $(\mathbf{O}_{\nu})_{12}$ and $(\mathbf{O}_{\nu})_{22}$. In order to make clear our statement, let us take the following value for the lightest neutrino mass, $m_{
1}=0.001$,  and central values for the squared mass scales, therefore, one gets $\vert (\mathbf{O}_{\nu})_{13}\vert\approx 0.023$,  $\vert (\mathbf{O}_{\nu})_{23}\vert\approx 0.35$, $\vert (\mathbf{O}_{\nu})_{12}\vert\approx 0.32$ and $\vert (\mathbf{O}_{\nu})_{22}\vert\approx 0.84$

Having fixed those particular values for the entries that take places in the mixing angles formulas, the reactor, atmospheric and solar angles values depend on the values of $\theta_{e}$ parameter and the CP-violating phase, $\eta_{\nu}$ (see Eq. (\ref{mixan})). Let us focus in the formula for the reactor angle and its corresponding central value $\sin^{2}{\theta_{13}}\approx 0.022$: 
(a) if $\sin{\theta_{e}}\sim \mathcal{O}(1)$, then  $\eta_{\nu}$ must have to be $0$ or $2\pi$ to cancel each other the involved terms in the reactor angle. In this scenario, the reactor angle is quite large ($\sin^{2}{\theta}_{13}\approx0.09$) in comparison to the experimental values; the atmospheric ($\sin^{2}{\theta}_{23}\approx0.033$) and solar $\sin^{2}{\theta}_{12}\approx0.9$ angles are tiny and large, respectively; (b) if $\theta_{e}=\pi/4$, in similar way to the above case, the phase $\eta_{\nu}$ must have to be $0$ or $2\pi$, so that, the reactor angle value decreases ($\sin^{2}{\theta}_{13}\approx0.05$), the atmospheric ($\sin^{2}{\theta}_{13}\approx0.075$)  and solar ($\sin^{2}{\theta}_{13}\approx0.72$)  angles increases and decreases, respectively;  (c) if $\sin{\theta}_{e}\approx \mathcal{O} (0.35)$ the reactor and atmospheric angles are below the allowed experimental bound ($\sin^{2}{\theta}_{13}\approx0.01$ and $\sin^{2}{\theta}_{23}\approx0.11$), the solar angles is within the allowed region ($\sin^{2}{\theta}_{12}\approx0.36$), these particular values for the mixing angles are reached by considering the $0$ or $2\pi$ for the CP-violating phase, $\eta_{\nu}$.

\item Case II: $\tilde{a}_{\nu}\approx\tilde{m}_{1}$. In this case, we have
	\begin{eqnarray}
	(\mathbf{O}_{\nu})_{13}&\approx& 0, \qquad
	(\mathbf{O}_{\nu})_{23}\approx\sqrt{\tilde{m}_{2}\left(1+\frac{\tilde{m}_{1}}{\tilde{m}_{2}}-\tilde{m}_{2}\right)},\nn\\
	(\mathbf{O}_{\nu})_{12}&\approx& 0,\qquad
	(\mathbf{O}_{\nu})_{22}\approx\sqrt{1-\tilde{m}_{2}}
	\end{eqnarray}
	
In this limit, the mixing angle formulas (see Eq. (\ref{mixan})) do not depend on the CP-violating phase, $\eta_{\nu}$, and there is no too much freedom to accommodate simultaneously the central values for mixing angles. To be more explicit, let us consider  $\tilde{m}_{1}=0.001$, then $\vert (\mathbf{O}_{\nu})_{23}\vert\approx 0.4$ and $\vert (\mathbf{O}_{\nu})_{22}\vert\approx 0.9$. Having calculated these values, with $\sin{\theta_{e}}=0.35$, the reactor angles can be in the allowed region ($\sin^{2}{\theta}_{13}\approx0.020$) but the atmospheric and solar angles are below their allowed bounds.

\end{itemize}

As main conclusion of this brief analysis, for the normal hierarchy, we get that $\tilde{a}_{\nu}\neq0$ and $\tilde{a}_{\nu}\neq \tilde{m}_{1}\neq 0.001$. Besides, we emphasize that the mixing angles are sensitive to the lightest neutrino mass so that drastic changes in the values are expected with a different value for the lightest neutrino mass.

\item [Inverted Hierarchy] ($1>\tilde{m}_{1}>\tilde{m}_{3}>\tilde{a}_{\nu}>0$). 

For this ordering, from Eq. (\ref{massfix}), we obtain the normalized neutrino masses
\begin{equation}
\tilde{m}_{1}=\sqrt{\frac{\Delta m^{2
		}_{23}+m^{2}_{3}-\Delta m^{2
	}_{21}}{\Delta m^{2
		}_{23}+m^{2}_{3}}},\qquad \tilde{m}_{3}=\frac{m_{3}}{\sqrt{\Delta m^{2
		}_{23}+m^{2}_{3}}}, \qquad \frac{\tilde{m}_{3}}{\tilde{m}_{1}}=\frac{m_{3}}{\sqrt{\Delta m^{2
		}_{23}+m^{2}_{3}-\Delta m^{2
	}_{21}}}.
\end{equation}

\begin{itemize}
	\item Case I: $\tilde{a}_{\nu}\approx 0$. For this case, we have the following
	\begin{eqnarray}
	(\mathbf{O}_{\nu})_{13}&\approx& \sqrt{1-\frac{\tilde{m}_{3}}{\tilde{m}_{1}}}, \qquad
	(\mathbf{O}_{\nu})_{23}\approx\sqrt{\frac{\tilde{m}_{3}}{\tilde{m}_{1}} \left(1-\frac{\tilde{m}_{3}}{\tilde{m}_{1}}-\tilde{m}_{1}\right)},\nn\\
	(\mathbf{O}_{\nu})_{12}&\approx&\tilde{m}_{1} \sqrt{\frac{\tilde{m}_{3}}{1-\tilde{m}^{2}_{1}}\left(1-\frac{\tilde{m}_{3}}{\tilde{m}_{1}}\right)},\qquad
	(\mathbf{O}_{\nu})_{22}\approx \sqrt{\frac{\tilde{m}_{1}}{1+\tilde{m}_{1}}\left(1-\frac{\tilde{m}_{3}}{\tilde{m}_{1}}\right)}
\end{eqnarray}

In the limit of strict inverted hierarchy, $m_{3}=0$, the following relations hold $\vert (\mathbf{O}_{\nu})_{13}\vert\gg \vert (\mathbf{O}_{\nu})_{23} \vert$	and $\vert (\mathbf{O}_{\nu})_{22}\vert> \vert (\mathbf{O}_{\nu})_{12} \vert$. For that reason, the reactor and solar angles will be controlled by the magnitude of the entry $(\mathbf{O}_{\nu})_{13}$; at the same time, the solar angle depends on the magnitude of the entry $(\mathbf{O}_{\nu})_{22}$. In order to fix some ideas on the values of the entries, let us consider the following value for the lightest neutrino mass, $m_{3}=0.0005$. Then, one obtains $\vert (\mathbf{O}_{\nu})_{13}\vert\approx 0.99$, $\vert (\mathbf{O}_{\nu})_{23}\vert\approx 0.007$, $\vert (\mathbf{O}_{\nu})_{12}\vert\approx 0.57$ and $\vert (\mathbf{O}_{\nu})_{22}\vert\approx 0.7$; given these values and according to the formulas for the mixing angles, we can not accommodate simultaneously the reactor and atmospheric angles since $\cos{\theta_{e}}\ll1$ is required by obtaining $\sin^{2}{\theta}_{13}\approx 0.022$ but this makes too small the atmospheric angle or vice versa. To be more explicit, let us consider $\sin{\theta}_{e}\approx 0.99$, then, we obtain $\sin^{2}{\theta}_{13}\approx 0.017$, $\sin^{2}{\theta}_{23}\approx 0.98$ and $\sin^{2}{\theta}_{12}\approx 0.37$; in this case the CP- violating phase is taken as $\eta_{\nu}=0$. If the value for the CP phase changed ($\eta_{\nu}=\pi$), the above mixing angle values would chance; the solar angle value is modified drastically from $0.37$ to $0.62$.

\item Case II: $\tilde{a}_{\nu}\approx\tilde{m}_{3}$. In this limit, we obtain
\begin{eqnarray}
(\mathbf{O}_{\nu})_{13}&\approx& 1, \qquad
(\mathbf{O}_{\nu})_{23}\approx 0,\nn\\
(\mathbf{O}_{\nu})_{12}&\approx& 0,\qquad
(\mathbf{O}_{\nu})_{22}\approx \sqrt{\frac{\tilde{m}_{1}}{1+\tilde{m}_{1}}\left(1+\frac{\tilde{m}_{3}}{\tilde{m}_{1}}\right)}
\end{eqnarray}

From Eq.(\ref{mixan}), one can observe that the CP-violating phase is irrelevant for these observables. In addition, in similar way to the above case, the reactor and the atmospheric angles can not accommodate simultaneously due to these depends strongly on the value the $\theta_{e}$ parameter. In order to get the central value for the reactor angle, one needs that $\sin{\theta}_{e}\approx \mathcal{O} (1)$ as consequence the atmospheric angle turns out being so large. Along with this, the solar angle also is large. 
\end{itemize}

As result of this brief analysis for the inverted hierarchy, one might conclude that this mass pattern is favored less than the normal hierarchy. In the next section, a numerical study will be carried out to find a set of free parameters that fits the observables.
\end{description}

\subsubsection{Numerical study}

In this sector, the numerical analysis  consists of scattering plots for constraining the allowed region for each free parameters. Notice that in this sector, as was pointed out, there are five free parameters that take place in the PMNS mixing matrix, however, only three play an important role in the reactor, solar and atmospheric angles namely $\vert b_{e}\vert$, $\vert a_{\nu}\vert$ and one effective CP violating phase, $\eta_{\nu}$. Actually, the lightest neutrino masses will be considered as free parameters then, at the end of the day, we have to constraint four parameters in the magnitude of the mixing values.

\begin{table}[th]
	\begin{center}
		\begin{tabular}{|c|c|c|}
			\hline\hline
			Observable & Experimental value \\ \hline
			$m_{e}(MeV)$ & \quad $0.5_{-0.45}^{+0.56}$ \\ \hline
			$m_{\mu}(MeV)$ &  \quad $106\pm 86$ \\ \hline
			$m_{\tau}(GeV)$ &  \quad $1116\pm 0.6\pm 0.9$ \\ \hline
			$\frac{\Delta m^{2}_{21}}{10^{-5}~eV^{2}}$ & \quad $7.40_{-0.20}^{+0.21}$ \\ \hline
			$\frac{\Delta m^{2}_{31}}{10^{-3}~eV^{2}}$ & \quad $2.494._{-0.031}^{+0.033}$ \\ \hline
			$\frac{\Delta m^{2}_{23}}{10^{-3}~eV^{2}}$ & \quad $2.465_{-0.031}^{+0.032}$ \\ \hline
			$\sin^{2}{\theta}_{12}$ & \quad $0.307^{+0.013}_{-0.012}$ \\ \hline
			$\sin^{2}{\theta}_{23}$ &  \quad $0.538^{+0.033}_{-0.069} $ ($0.554^{+0.023}_{-0.033}$)\\ \hline
			$\sin^{2}{\theta}_{13}$ & \quad $0.02206\pm0.00075$ ($0.02227\pm0.00074$)\\ \hline
			$\delta_{CP}/^{\circ}$ & \quad $ 243^{+43}_{-31}$ ($ 278^{+26}_{-29}$) \\ \hline\hline
		\end{tabular}%
	\end{center}
	\caption{ Experimental values of the lepton masses and PMNS parameters \cite{Patrignani:2016xqp}.}
	\label{Tabpdg}
\end{table}

Summarizing, the mixing angles are functions of the following parameters
\begin{eqnarray}\label{mixan2}
\sin{\theta}_{13}&=&\sin{\theta}_{13}\left(\vert b_{e}\vert, \vert a_{\nu}\vert, \eta_{\nu}, m_{j} \right)\nn\\
\sin{\theta}_{23}&=&\sin{\theta}_{23}\left(\vert b_{e}\vert, \vert a_{\nu}\vert, \eta_{\nu}, m_{j} \right) \nn\\
\sin{\theta}_{12}&=&\sin{\theta}_{12}\left(\vert b_{e}\vert, \vert a_{\nu}\vert, \eta_{\nu}, m_{j} \right)
\end{eqnarray}
where  $j=1, 3$ stands for the normal and inverted hierarchy, respectively. Along with this, as we already did it, two neutrino masses can be fixed in terms of the squared mass scales and the lightest neutrino mass. 

Having done that, in the scattering plots, we will vary the free parameters in such a way those satisfy their respective constraints. For the lightest neutrino mass, in the normal (inverted) case, we have $1>\tilde{m}_{2}>\tilde{m}_{1}>\tilde{a}_{\nu}>0$  ($1>\tilde{m}_{1}>\tilde{m}_{3}>\tilde{a}_{\nu}>0$); along with this, for each hierarchy, the lightest mass let vary in the region $0-0.9~eV$. At the same time, the effective CP-violating phase $2\pi\geq \eta_{\nu}\geq0$ and the charged lepton parameter $m_{\mu}> \vert b_{e}\vert>m_{e}$. Then, we demand that the our theoretical expressions for the observables satisfy the experimental bounds up to $3\sigma$, then, this allows us to scan the allowed regions for the free parameters that fit quite well the experimental results. As a model prediction, the CP-violating phases is fitted.

\begin{figure}[ht!]
\centering
\includegraphics[scale=0.45]{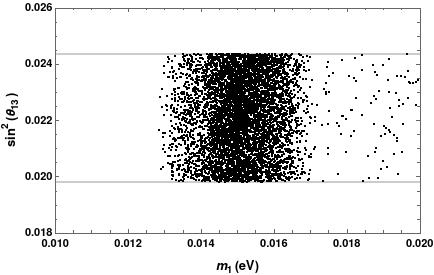}
\hspace{1mm}\includegraphics[scale=0.45]{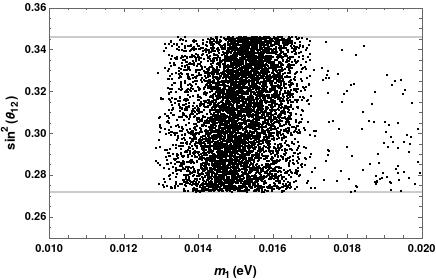}\\
\includegraphics[scale=0.45]{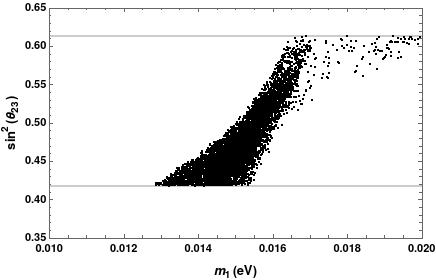}\hspace{1mm}\includegraphics[scale=0.45]{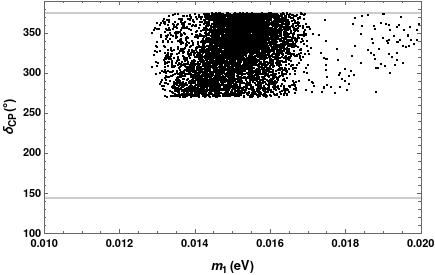}
\caption{From left to right: the reactor, solar, atmospheric angles and CP phase versus the lightest neutrino mass. The thick line stands for $3~\sigma$ of C. L.}\label{f1}
\end{figure}

As we can observe, in the Fig. (\ref{f1}), there is an allowed region ($0.013-0.017$~eV) for the lightest neutrino mass that accommodates very well the observables. 

\begin{figure}[h!]\centering
	\includegraphics[scale=0.45]{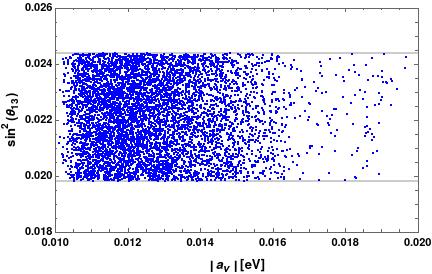}
	\hspace{1mm}\includegraphics[scale=0.45]{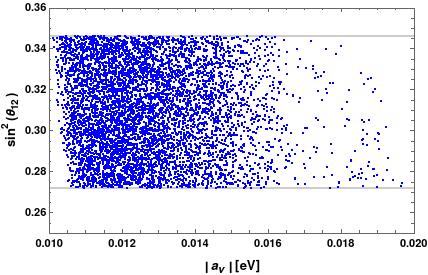}\\
	\includegraphics[scale=0.45]{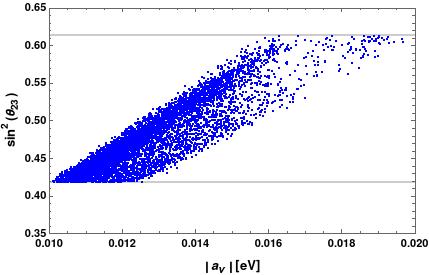}\hspace{1mm}\includegraphics[scale=0.45]{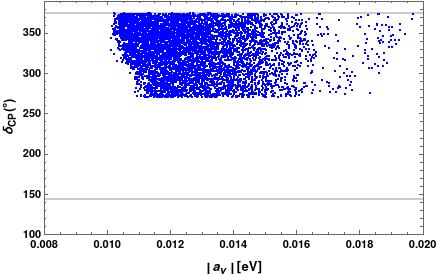}
\caption{From left to right: the reactor, solar, atmospheric angles and CP phase versus the $\vert a_{\nu}\vert$ parameter. The thick line stands for $3~\sigma$ of C. L.}\label{f2}
\end{figure}

In the Fig. (\ref{f2}), the observables are shown as function of the free parameter $\vert a_{\nu}\vert$ ($\tilde{a}_{\nu}=\vert a_{\nu}\vert/m_{3}$) which satisfies the constraint $1>\tilde{m}_{2}>\tilde{m}_{1}>\tilde{a}_{\nu}>0$. According to our numerical analysis, the observables are fitted with $\vert a_{\nu}\vert\approx m_{1}$. This case corresponds to the second scenario, for the normal hierarchy. However, in the previous analytical study, we just considered the particular value for the lightest neutrino mass, $\tilde{m}_{1}=0.001$ which is a limit case for strict normal hierarchy. Due to the numerical results, ${m}_{1}$ is not tiny and there is a region for the free parameter, $\vert a_{\nu}\vert$, where the mixing angles are in good agreement with the experimental data. 

\begin{figure}[h!]\centering
	\includegraphics[scale=0.45]{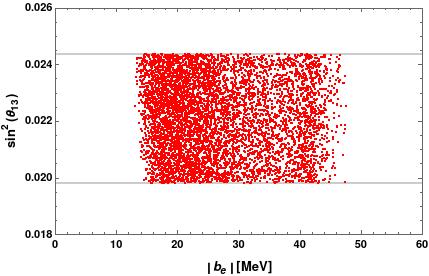}
	\hspace{1mm}\includegraphics[scale=0.45]{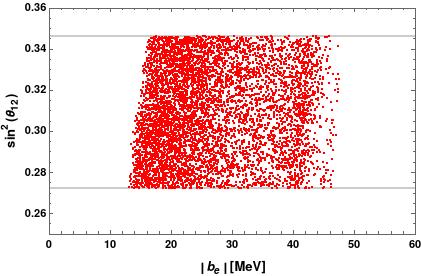}\\
	\includegraphics[scale=0.45]{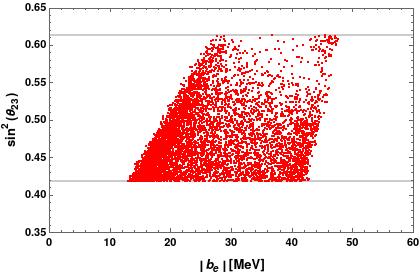}\hspace{1mm}\includegraphics[scale=0.45]{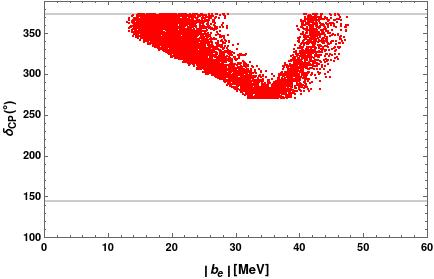}
\caption{From left to right: the reactor, solar, atmospheric angles and CP phase versus the $\vert b_{e}\vert$ parameter. The thick line stands for $3~\sigma$ of C. L.}\label{f3}
\end{figure}

As we already commented, for the normal hierarchy, the favored scenario corresponds with $\vert a_{\nu}\vert\approx m_{1}$. In this case, the mixing angles and the CP-violating phase as function of the free parameter, $\vert b_{e}\vert$, are shown in the Fig. (\ref{f3}). Let us remind you that $m_{\mu}>\vert b_{e}\vert>m_{e}$, then the numerical study allows to find a window of values ($10-45$~MeV) where the observables are fitted.

\begin{figure}[h!]\centering
	\includegraphics[scale=0.45]{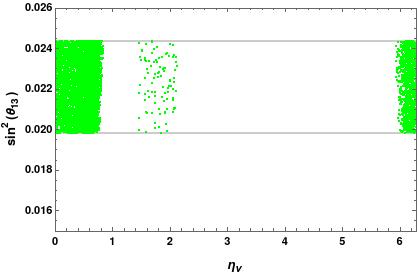}
	\hspace{1mm}\includegraphics[scale=0.45]{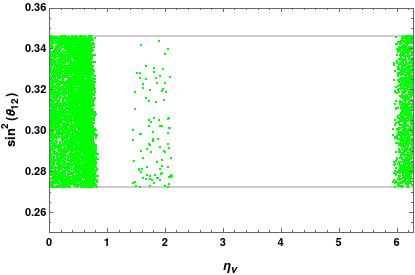}\\
	\includegraphics[scale=0.45]{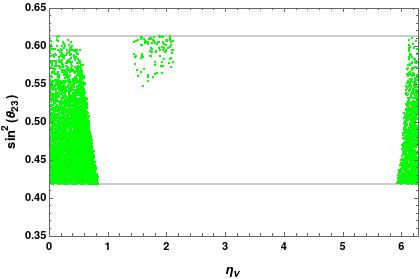}\hspace{1mm}\includegraphics[scale=0.45]{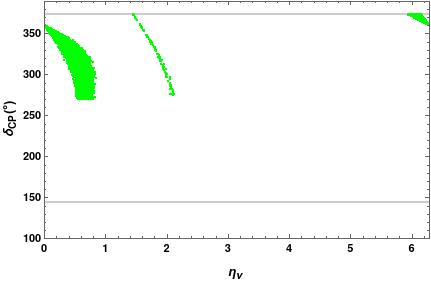}
	\caption{From left to right: the reactor, solar, atmospheric angles and CP phase versus the effective phase, $\eta_{\nu}$, parameter. The thick line stands for $3~\sigma$ of C. L.}\label{f4}
\end{figure}

The last plot, Fig.(\ref{f4}), shows the observables as function of the only CP phase, $\eta_{\nu}$. As was mentioned in the analytical study, for the second scenario ($\vert a_{\nu}\vert\approx m_{1}$), the mixing angles were favored by considering $\eta_{\nu}=0$ or $\eta_{\nu}=2\pi$. Our statement gets support by the scattering plot, actually, the best region of values, that accommodate the observables, is close to $\eta_{\nu}$.

Let us focus in the inverted hierarchy, as it was shown, the analytical study reveled that the mixing angles were not accommodate simultaneously. Besides, the numerical study provides a strong support to rule out the inverted hierarchy as we will see next.

In similar way to the normal hierarchy, we will vary the free parameters in such a way those satisfy the constraints. For the lightest neutrino mass, we have   $1>\tilde{m}_{1}>\tilde{m}_{3}>\tilde{a}_{\nu}>0$. Along with this, for each hierarchy, the lightest mass let vary in the region $0-0.9~eV$. At the same time, the effective CP-violating phase $2\pi\geq \eta_{\nu}\geq0$ and the charged lepton parameter $m_{\mu}> \vert b_{e}\vert>m_{e}$. Having commented that, in the present ordering, we demand that the our theoretical expression for the solar angle satisfies the experimental bound up to $3\sigma$, then, this allows us to scan the allowed regions for the free parameters that fit quite well the experimental results. 

\begin{figure}[h!]\centering
	\includegraphics[scale=0.45]{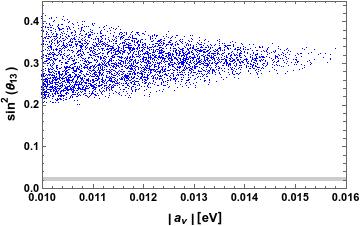}
	\hspace{1mm}\includegraphics[scale=0.45]{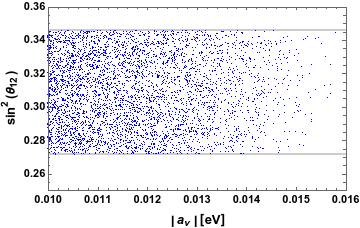}\\
	\includegraphics[scale=0.45]{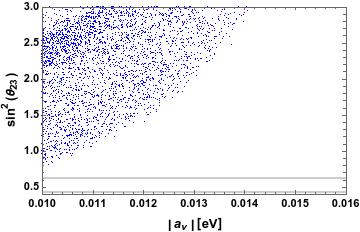}\hspace{1mm}\includegraphics[scale=0.45]{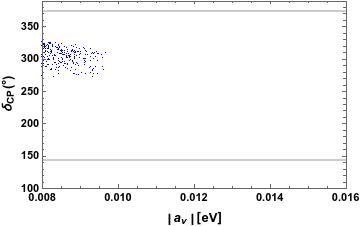}
	\caption{From left to right: the reactor, solar, atmospheric angles and CP phase versus the $\vert a_{\nu}\vert$ parameter. The thick line stands for $3~\sigma$ of C. L.}\label{f5}
\end{figure}

The main finding is the following: the mixing angles are not accommodated in good agreement with the experimental results. In the figure (\ref{f5}), we see the mixing angles as function of the free parameter $\vert a_{\nu}\vert$, in this case, one can not find a set of values for the $\vert a_{\nu}\vert$ parameter that fits the observables. Thus,  
the solar angles is fixed at $3\sigma$ whereas the reactor and the atmospheric angles are beyond their allowed regions.

\section{Lepton violation process: $\mu\rightarrow e \gamma$}

As model prediction, we have calculated the branching ratio for the lepton flavor violation process $\mu\rightarrow e \gamma$ \cite{Akeroyd:2009nu, Lindner:2016bgg} that is mediated by the doubly ($\Delta^{++}$) and singly ($\Delta^{+}$) charged scalars that come from the Higgs triplet (see Eq.(\ref{scal})). The branching ratio \cite{Akeroyd:2009nu} is given by 

\begin{equation}
\textrm{BR}(\mu\rightarrow e \gamma)\approx 4.5\times 10^{-3}\left(\frac{1}{\sqrt{2}v_{\Delta}\lambda}\right)^{4}\left| \left( \mathbf{V}^{\ast}\hat{\mathbf{M}}^{\dagger}_{\nu}\hat{\mathbf{M}}_{\nu} \mathbf{V}^{T}\right)_{e\mu}\right|^{2}\left(\frac{200~GeV}{m_{\Delta^{++}}}\right)^{4}
\end{equation}
where $m_{\Delta^{+}}=m_{\Delta^{++}}\equiv m_{\Delta}$ has been assumed in the previous result. Besides, $\mathbf{V}$ stands for the PMNS mixing matrix.

The branching ratio depends on the PMNS mixing parameters, the single and doubly charged scalars; along with this, the vev of the Higgs triplet takes place. In here, we use the following regions $80~GeV< m_{\Delta}$ and $v_{\Delta}<5~GeV$~\cite{CarcamoHernandez:2018djj}; the PMNS mixing parameters have been already constrained in the previous section, to be more explicit, we use the following regions: $0.01~eV<m_{1}<0.02~eV$, $10~MeV<\vert b_{e}\vert<60~MeV$, $0.01~eV<\vert a_{\nu}\vert<0.02~eV$ and $0< \eta_{\nu}<\pi/2$. We have to point out that $\vert a_{\nu}\vert<m_{1}$ and the effective CP-violating phase, $\eta_{\nu}$, can vary in the interval $[0,2\pi]$ but the numerical study showed the favored region lies in $[0,\pi/2]$ according the figure (\ref{f4}).

\begin{figure}[h!]\centering
	\includegraphics[scale=0.45]{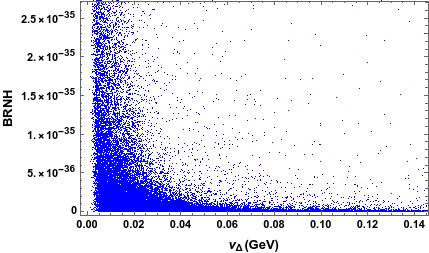}
	\hspace{10mm}\includegraphics[scale=0.45]{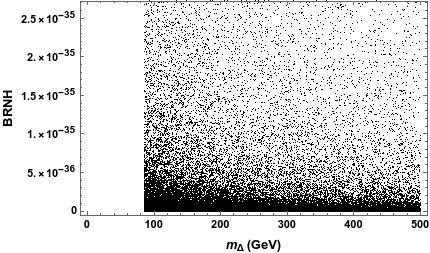}
	\caption{From left to right: BR($\mu\rightarrow e\gamma$) versus the $v_{\Delta}$ and $m_{\Delta}$ parameter. The thick line stands for $3~\sigma$ of C. L.}\label{f6}
\end{figure}

In the Fig. (\ref{f6}), the predicted region is shown
for the branching ratio as function of the vev of the Higgs triplet and the mass of the singly and doubly charged scalars. Our model predicted a region, $\textrm{BR}(\mu \rightarrow e\gamma)\approx 10^{-35}$, that is too much below of the experimental bound $\textrm{BR}(\mu \rightarrow e\gamma)\approx 4.2\times10^{-13}$.

\section{Conclusions}

A flavored non-renormalizable model has been introduced with the main purpose of accommodating simultaneously the mixings for the quark and lepton sectors. We have to pay a high price to do so, this is, apart from two extra Higgs doublets and one triplet, flavon scalars were needed to get desirable mass textures in the fermion mass matrices. Along with this, strong assumption were made in order to use the minimal discrete symmetries (one $\mathbf{Z}_{2}$ for both sectors).

Our findings are the following: in the quark sector, the hierarchy, among the quark masses, takes relevancy to obtain the mixing matrix since that the NNI textures are behind the mixings so that the CKM mixings are fit quite well. In the lepton sector, the type II see-saw mechanism is responsible to explain tiny neutrino masses; due to the flavor symmetry, the neutrino mass matrix possesses a kind of extended Fritzsch textures which contribute mainly to the PMNS mixing matrix. The model favors the normal hierarchy which is preferred by the current oscillations data. Then, it was possible to find a set of values for the four free parameters that accommodate the observables in good agreement with the experimental results.

Although the model is naive and too much elaborated, this framework encourage us study the $\mathbf{S}_{3}$ flavor symmetry since its irreducible representations may be a key to understand the hierarchy among the fundamental particles.

\section*{Acknowledgements}
The authors want to thank Myriam Mondrag\'on for her valuable comments on the manuscript. Garc\'ia-Aguilar appreciates the facilities given by the IPN through the SIP project number 20201313. JCGI thanks  Valentina A. and A. Emiliano  G\'omez Nabor for  sharing great moments and experiences during this long time. This work was partially supported by Project 20202024, the Mexican grants 237004 and PAPIIT IN111518.

\appendix
\section{$\mathbf{S}_{3}$ flavour symmetry}
The non-Abelian group ${\bf S}_{3}$ is the permutation group of three objects \cite{Ishimori:2010au} and this has three irreducible representations: two 1-dimensional, ${\bf 1}_{S}$ and ${\bf 1}_{A}$, and one 2-dimensional representation, ${\bf 2}$. We list the multiplication rules among them:

\begin{eqnarray}\label{rules}
{\bf 1}_{S}&\otimes& {\bf 1}_{S}={\bf 1}_{S}, \qquad {\bf 1}_{S}\otimes {\bf 1}_{S}={\bf 1}_{S},\qquad {\bf 1}_{S}\otimes {\bf 1}_{A}={\bf 1}_{A}\nonumber\\
{\bf 1}_{A}&\otimes& {\bf 1}_{A}={\bf 1}_{S},\qquad {\bf 1}_{S}\otimes {\bf 2}={\bf 2},\qquad {\bf 1}_{A}\otimes {\bf 2}={\bf 2}\nonumber\\
\begin{pmatrix}
a_{1} \\ 
a_{2}
\end{pmatrix}_{{\bf 2}}
&\otimes&
\begin{pmatrix}
b_{1} \\ 
b_{2}
\end{pmatrix}_{{\bf 2}}=
\left(a_{1}b_{1}+a_{2}b_{2}\right)_{{\bf 1}_{S}} \oplus  \left(a_{1}b_{2}-a_{2}b_{1}\right)_{{\bf 1}_{A}} \oplus	
\begin{pmatrix}
a_{1}b_{2}+a_{2}b_{1} \\ 
a_{1}b_{1}-a_{2}b_{2}
\end{pmatrix}_{{\bf 2}}. 
\end{eqnarray}

\bibliographystyle{bib_style_T1}
\bibliography{references.bib}

\end{document}